\documentclass[final,1p,times]{elsarticle}

\usepackage{amssymb}
\usepackage{amsmath}
\usepackage{amsthm}

\journal{Chaos, Solitons and Fractals}

\begin{document}

\begin{frontmatter}



\title{Phase space noncommutativity, power-law inflation and quantum cosmology}


\author[1,2]{S. M. M. Rasouli}
\author[1]{Jo\~{a}o Marto}

\address[1]{Departamento de F\'{i}sica,
Centro de Matem\'{a}tica e Aplica\c{c}\~{o}es (CMA-UBI),
Universidade da Beira Interior,
Rua Marqu\^{e}s d'Avila
e Bolama, 6200-001 Covilh\~{a}, Portugal}

\begin{abstract}
Considering an arbitrary dimensional FLRW universe in the framework of a generalized S\'{a}ez--Ballester (SB) theory,
we establish a noncommutative (NC) cosmological model. We concentrate on the predictions of NC model and compare them with their commutative counterparts in both the classical and quantum regimes. For the classic case, taking a very small NC parameter, we apply two different methods to analyze the model features. First, we show through numerical analysis that our NC model is a successful inflationary model capable of overcoming the graceful exit and horizon problems. Furthermore, the NC traces are visible the late time, which supports the UV/IR mixing characteristic of the NC models. In the second method, we show that our NC model can correspond to the previously developed NC inflationary models. In the commutative quantum case, we obtain an exact wave function and then use the WKB approximation to show that the solutions of the corresponding classical regime are recovered. Finally, with regard to the NC quantum level, we focus on the special case for which we show that a constant of motion exists. The latter helps us to conveniently transform the corresponding complicated NC-WDW equation into an ordinary differential equation, which can be easily solved numerically for the general case
or analytically for some special cases. The resultant solutions show a damping behavior in the wave function associated with the proposed NC model, which may be important in determining the viable initial states for the very early universe.

\end{abstract}

\begin{keyword}
S\'{a}ez--Ballester theory \sep Noncommutativity  \sep Inflation \sep Quantum cosmology \sep WDW equation
 \sep FLRW Cosmology \sep UV/IR mixing \sep WKB approximation

\end{keyword}

\end{frontmatter}

\section{Introduction}
\label{SecI}
\indent

Models based on noncommutative (NC) theories are still of interest.
The main motivation for employing noncommutativity of the canonical type may be traced back to the investigations that establish its relationship with string and M theory \cite{SW99}. In addition, the works in semi--classical gravity \cite{DFR94} could be related to NC field theories.
Furthermore, the application of NC theories allows for the consideration of several fascinating aspects, such as physics at extremely small distances \cite{S03}, non-locality, IR/UV mixing \cite{MRS00}, Lorentz violation \cite{CHKLO01}, the NC Landau problem \cite{H02} and the quantum Hall effect \cite{B86}; as other interesting applications of the noncommutativity in physics, see also \cite{AM15,AM23,AM24}.

Many NC scenarios have been established as cosmological applications, see for example Newtonian cosmology \cite{RS03}, quantum cosmology \cite{GOR02}, perturbation cosmology and NC inflation cosmology \cite{BH02}.
Let us be more precise. NC cosmological models have the ability to solve some of the most important outstanding problems in cosmology by opening up new pathways for studying the fundamental nature of the cosmos, see, for instance \cite{RFK11, RM14, JRM14, RZMM14, RZJM16, RM16, OM21, PG22, Addazi22, LMP23, KAS23} and references therein.
Modified dispersion relations, NC inflation, dark matter and dark energy, the cosmic microwave background, and quantum gravity effects are just a few of the relevant issues that the NC cosmological scenarios can predict appropriately, see for instance \cite{SRT24,AGM22} and references therein.

Quantum effects are expected to play a key role in the cosmic evolution of the universe. Unifying GR with quantum mechanics has been a outstanding problem, for which  NC geometry could provide insights and aid in the development of a full theory of quantum gravity. Among the various candidates for quantum gravity, many models of NC gravity have been proposed \cite{BH02,GOR02}. Most NC models assume a homogeneous universe where the metric (as well as the applied scalar field) depends only on time. The resulted model is a finite-dimensional configuration space  called a minisuperspace.

The main objective of this work is to investigate of the traces of noncommutativity in early times in (semi-) classical and quantum regimes. Concretely, we propose a dynamical noncommutativity for a cosmological model in which we consider a deformation of minisuperspace rather than a deformation of spacetime. Let us be more precise. In the classical regime (for both the commutative and the NC models), as well as in the commutative quantum scenario, we will consider a specific potential, so an extended version of the standard power-law inflationary model provided in Ref. \cite{LM85} is elaborated.
Subsequently, in the NC classical scenario, we will demonstrate not only that our model has the characteristics of a successful inflationary model, but also, using another method for solving differential equations, that our model corresponds to an NC inflationary model established in Ref. \cite{R11}.
Finally, to explore solutions for our quantum NC model, we focus on a specific case where the potential takes a constant value and the chosen dynamic noncommutativity is assumed to be the simplest case. To present our analysis for this case we
use Seiberg-Witten (SW) map, which helps convert
the NC system into an appropriately modified commutative one.

The paper is organized as follows.
In the next section, considering the FLRW metric, we proposed an NC cosmological model in the context of the S\'{a}ez--Ballester (SB) theory in arbitrary dimensions. In sections \ref{SecIII} and \ref{SecIV}, we study the commutative and NC SB cosmology in classical regime and compare the corresponding results. In sections \ref{SecV} and \ref{SecVI}, we investigate the quantum version of the models mentioned. Due to the complexity of the field equations for the NC quantum model, we restrict ourselves to a particular case.
In Section \ref{Concl}, we provide an overview of the models presented and highlight some important findings.


\section{Noncommutative SB cosmology}
\label{SecII}

Let us consider the $D$-dimensional
Friedmann--Lema\^{i}tre--Robertson--Walker (FLRW) metric as the background geometry:
\begin{eqnarray}
\label{FRW-met-1}
ds^{2}=-N(t) dt^{2}+a^{2}(t)\left(\frac{dr^2}{1-\mathcal K r^2}+r^2d\Omega_{_{D-2}}^2\right),
\end{eqnarray}
where $\mathcal K=-1,0,1$; $N(t)$ and $a(t)$ denote the lapse function and the scale factor, respectively; $d\Omega_{_{D-2}}^2=d\theta_1^2+sin^2\theta_1d\theta_2^2+...+sin^2\theta_1
...sin^2\theta_{D-3}d\theta_{D-2}^2$ for $D\geq3$.

Moreover, instead of the conventional
S\'{a}ez--Ballester (SB) theory \cite{SB86}, we consider a generalized
version of it, which contains a scalar potential, in $D$ dimensions
as an underlying gravitational framework:
\begin{eqnarray}\label{SB-action}
 {\cal S}_{_{\rm SB}}^{^{(D)}}=\int d^{^{\,D}}\!x  {\cal L}=\int {\rm d}t d^{^{D-1}}\!x \sqrt{-g}\,
 \Big[R^{^{(D)}}-{\cal W}\phi^n\, g^{\alpha\beta}\,({\nabla}_\alpha\phi)({\nabla}_\beta\phi)
 -V(\phi)\,-2\chi \varrho(a)\Big],
\end{eqnarray}
where ${\cal W}$ and $n$ are the parameters of the model; $\chi\equiv8\pi G$; the homogeneous scalar filed $\phi$ is minimally coupled to the Ricci scalar; the Greek indices run from zero to $D-1$; the covariant derivative is denoted by $\nabla$;
$g$ and $R^{^{(D)}}$ are the determinant and Ricci scalar associated with the
$D$-dimensional metric $g_{\alpha\beta}$, respectively and we used the units where $c=1=\hbar$.
From now on, we will consider only the spatially flat FLRW universe.

The Ricci scalar corresponding to the spatially flat FLRW metric \eqref{FRW-met-1} is given by
\begin{equation}\label{FRW-Ricci}
R^{^{(D)}}=\frac{2(D-1)}{N^2}\left[\frac{\ddot{a}}{a}+\frac{D-2}{2}\left(\frac{\dot{a}}{a}\right)^2 -\left(\frac{\dot{N}}{N}\right)\left(\frac{\dot{a}}{a}\right)\right].
\end{equation}
Substituting $R^{^{(D)}}$ from \eqref{FRW-Ricci} into \eqref{SB-action}, we obtain
 \begin{equation}\label{FRW-action-2}
{\cal L}=-Na^{D-1}
\left[\frac{(D-1)(D-2)}{N^2}\left(\frac{\dot{a}}{a}\right)^2
-\frac{{\cal W}\phi^n}{N^2}\dot{\phi }^2
+V(\phi )+2\chi \varrho(a)\right],
\end{equation}
where $\dot{A}\equiv dA/dt$.

The conjugate momenta associated with the scale factor and scalar field are:
\begin{eqnarray}
\label{momenta}
\Pi _a&=& -2(D-1)(D-2){N}^{-1}a^{D-3}\dot{a},\\\nonumber\\
\Pi _{\phi } &=&2{\cal W}{N}^{-1}a^{D-1}\phi^n\dot{\phi }.
\end{eqnarray}
$m=4$
From using the standard relationship between the Lagrangian and the classical
Hamiltonian $\mathcal{H}$, it is straightforward to show that
 \begin{eqnarray}
\label{Hamil1}
\mathcal{H}= -\frac{{\Pi ^2_a}}{4(D-1)(D-2)a^{D-3}}
+\frac{\Pi ^2_{\phi }}{4{\cal W}a^{D-1}\phi^n}
+a^{D-1}\left[V(\phi )+2\chi \varrho(a)\right],
 \end{eqnarray}
where, for the sake of technical simplicity, we used the particular
gauge $N=1$, which is only relevant for the classic level.
It should be noted that, at the quantum level, the lapse function does not appear in the field equations.

Let us propose a NC cosmological model considering a deformed Poisson algebra as
\begin{eqnarray}\label{deformed}
\{a,\Pi_a\}=1=\{\phi,\Pi_{\phi }\}, \hspace{5mm}
\{\Pi_a,\Pi_{\phi }\}=\theta\phi^{\kappa},
\end{eqnarray}
where the other Poisson brackets vanish, $\theta={\rm constant}$ is the NC parameter and
\begin{eqnarray}\label{alpha}
\kappa\equiv\frac{D(n+2)}{D-2}-1,
\end{eqnarray}
which is proposed based on the dimensional analysis; see
also Refs. \cite{RFK11,BBDP08, MPS11} for more information on
why it is interesting to propose noncommutativity between momenta.

Employing relations \eqref{deformed} and
\begin{eqnarray}\label{calc}
\left\{\Pi_{a },f(\Pi_{a },{\Pi}_{\phi})\right\}=\theta\phi^{\kappa}\frac{\partial f}{\partial {\Pi}_{\phi}},\hspace{8mm}
\left\{{\Pi}_{\phi},f(\Pi_{a },{\Pi}_{\phi})\right\}=-\theta\phi^{\kappa}\frac{\partial f}{\partial \Pi_{a }},
\end{eqnarray}
where $f$ is an arbitrary function of the momenta.
The classical equations of motion for the phase
space variables $a$, $\phi$, $\Pi_{a }$ and $\Pi_{\phi }$ are given by
\begin{eqnarray}
\dot{a} & = &
-\frac{1}{2(D-1)(D-2)}a^{3-D}\Pi_{a},
\label{diff.eq1}\\\nonumber\\\nonumber
\dot\Pi_{a }\!\!\! & = &
-\frac{D-3}{4(D-1)(D-2)}a^{2-D}\Pi_{a}^{2}+\frac{D-1}{4{\cal W}}a^{-D}\phi^{-n}\Pi_{\phi}^{2}\\\nonumber\\\nonumber
&-&(D-1)a^{D-2}\left[V(\phi )+2\chi  \varrho(a)\right]
-2\chi  a^{D-1} \frac{d \varrho(a)}{da}+\frac{\theta }{2{\cal W}}a^{1-D}\phi^{\kappa-n}\Pi_{\phi},\\
\label{diff.eq2}\\\nonumber\\
\dot{\phi}\!\!\! & = & \frac{1}{2{\cal W}}a^{1-D}\phi^{-n}\Pi_{\phi},
\label{diff.eq3}\\\nonumber\\\nonumber
\dot{\Pi}_{\phi}& =&\frac{n}{4{\cal W}}a^{1-D}{\phi}^{-(n+1)}\Pi_{\phi}^2-a^{D-1}\frac{dV(\phi)}{d\phi}
+\frac{\theta}{2(D-1)(D-2)}a^{3-D}\phi^{\kappa}\Pi_{a}\\
\label{diff.eq4}
\end{eqnarray}
and
\begin{eqnarray}\label{constraint-asli}
\mathcal{H}=-\frac{{a^{3-D}\Pi ^2_a}}{4(D-1)(D-2)}
+\frac{a^{1-D}\phi^{-n}\Pi ^2_{\phi }}{4{\cal W}}
+a^{D-1}\left[V(\phi )+2\chi \varrho(a)\right]\approx0.
\end{eqnarray}
After some manipulations, we obtain the
classical equations of motion:
\begin{eqnarray}\label{asli1}
(D-1) H^{2}&=&\rho_{_{\phi}}+\frac{2\chi}{(D-2)}\varrho(a),\\\nonumber\\
\label{asli2}
2\frac{\ddot{a}}{a}+(D-3)H^{2}\!&=&\!-\left(p_{_{\phi}}+p_{_{\rm nc}}\right)+\frac{2\chi}{D-2}\left[\varrho(a)+\frac{a}{D-1}\frac{d\varrho(a)}{da}\right],\\\nonumber\\
\label{asli3}
\ddot{\phi}+(D-1)H\dot{\phi}\!&+&\!\frac{n}{2}{\phi}^{-1}\dot{\phi}^{2}+\frac{1}{2{\cal W}}{\phi}^{-n}\frac{dV(\phi)}{d\phi}+\frac{\theta  a^{2-D}H\phi^{\kappa-n}}{2{\cal W}} =0,
\end{eqnarray}
where
$H\equiv\dot{a}/a$; $\rho_{_{\phi}}$ and $p_{_{\phi}}$ denote the energy density and pressure
of the homogeneous scalar field:
\begin{eqnarray}\label{rho-tot}
\rho_{_{\phi}}&\equiv&\frac{1}{D-2}\left[{\cal W}{\phi}^{n}\dot{\phi}^{2}+V(\phi)\right],\\\nonumber\\
\label{p-tot}
p_{_{\phi}}&\equiv&\frac{1}{D-2}\left[{\cal W}{\phi}^{n}\dot{\phi}^{2}-V(\phi)\right].
\end{eqnarray}
Moreover, the quantity $p_{_{\rm nc}}$, which refers to an effective pressure
term arising due to the presence of the chosen dynamical deformation \eqref{deformed}, was defined as
\begin{eqnarray}\label{p-nc}
p_{_{\rm nc}}\equiv\frac{\theta}{(D-1)(D-2)}a^{2-D} \phi^{\kappa}\dot{\phi}.
\end{eqnarray}

Obtaining analytic solutions for the NC equations \eqref{asli1}-\eqref{asli3} seems very complicated, so in Section \ref{SecIV} we will present predictions for some physical quantities using numerical analysis. In addition, solving the NC-WDW equation for the corresponding quantum cosmological model is also not easy. Therefore, in Section \ref{SecVI}, we will only focus on a particular case.

In this work we are interested in studying the early universe where only the scalar field dominates the dynamics. Therefore, from now on we restrict ourselves to cosmological models that do not take ordinary matter into account.
In this case, it is easy to show that the following conservation equation is identically satisfied:
\begin{eqnarray}\label{cons-eq}
\Delta \equiv\dot{\rho}_{_{ \phi}}+(D-1)H\left(\rho_{_{ \phi}}+p_{_{ \phi}}+p_{_{\rm nc}}\right)=0.
\end{eqnarray}

Among the differential equations \eqref{asli1}-\eqref{asli3}
and \eqref{cons-eq}, only two of them are independent.
To specify these two, we would like to focus on the following equations:
\begin{eqnarray}\label{key-1}
H=\epsilon\sqrt{\frac{\rho_{_{\phi}}}{D-1}},
\end{eqnarray}
\begin{eqnarray}\label{key-2}
\frac{\dot{\rho}_{_{ \phi}}}{\sqrt{\rho_{_{\phi}}}}=-\frac{\sqrt{D-1}}{D-2}\left[2{\cal W}{\phi}^{n}\dot{\phi}^{2}+\frac{\theta}{D-1}a^{2-D} \phi^{\kappa}\dot{\phi}\right],
\end{eqnarray}
where $\epsilon \equiv\pm 1$.

Let us assume \footnote{
The justifications for choosing this type of potential are outlined in Ref.
\cite{RSM22}. Except for Section \ref{SecVI}, this potential will be used in all sections of this paper.}
\begin{eqnarray}\label{A-1}
V(\phi)=\left(\frac{2}{\Gamma}-{\cal W}\right)\phi^n\dot{\phi}^2,
\end{eqnarray}
where $\Gamma$ is a dimensionless constant.

By substituting $V(\phi)$ from \eqref{A-1} into
equations \eqref{rho-tot}, \eqref{key-1} and \eqref{key-2}, we obtain
\begin{align}\label{rho-sol1}
\rho_{_{\phi}}= & \: \frac{2{\phi}^{n}\dot{\phi}^{2}}{(D-2)\Gamma},\\\nonumber\\
\label{H-sol1}
H=& \: \epsilon h_0  \phi^{\frac{n}{2}}\dot{\phi},\\
\label{AE-1}
\frac{d}{dt}\left(\phi^n\dot{\phi}^2\right)=& \:
-\epsilon\sqrt{\frac{(D-1)\Gamma}{2(D-2)}}\left[2{\cal W}{\phi}^{n}\dot{\phi}^{2}+\frac{\theta}{(D-1)}
a^{2-D}\phi^{\kappa}\dot{\phi}\right] \phi^{\frac{n}{2}}\dot{\phi},
\end{align}
where we have used \begin{eqnarray}\label{eps}
\sqrt{\phi^n\dot{\phi}^2}=\epsilon \phi^{\frac{n}{2}}\dot{\phi},
  \hspace{15mm} h_0\equiv\sqrt{\frac{2}{(D-1)(D-2)\Gamma}}.
\end{eqnarray}
Equations \eqref{H-sol1} and \eqref{AE-1} are the independent equations associated with our proposed NC model for the chosen potential \eqref{A-1}.


\section{Commutative classical cosmology}
\label{SecIII}
In this case, substituting $\theta=0$ into
 equation \eqref{AE-1}, we obtain
\begin{eqnarray}\label{indep-1}
\left(\frac{1}{\phi^{\frac{3n}{2}}\dot{\phi}^3}\right)\frac{d}{dt}\left(\phi^n\dot{\phi}^2\right)=\epsilon \zeta,
\end{eqnarray}
where
\begin{eqnarray}\label{zeta}
\zeta\equiv -{\cal W}\sqrt{2\Gamma\left(\frac{D-1}{D-2}\right)}=-(D-1){\cal W}\Gamma h_0.
\end{eqnarray}

We can easily obtain a solution for \eqref{indep-1} as
\begin{eqnarray}\label{AES-1}
\phi^{\frac{n}{2}}\dot{\phi}=-\left(\frac{2\epsilon}{\zeta}\right)\frac{1}{t},
\end{eqnarray}
where the integration constant has been set equal to zero.

Equation \eqref{AES-1} gives
\begin{equation}\label{AES-2}
\phi(t)=\left \{
 \begin{array}{c}
\left[\phi_i^{\frac{n+2}{2}}-\frac{\epsilon(n+2)}{\zeta} \,\, {\rm ln}
\left(\frac{t}{t_i}\right)\right]^{\frac{2}{n+2}},
 \hspace{14mm} {\rm for}\hspace{7mm} n\neq-2,\\\\
 \phi_i\left(\frac{t}{t_i}\right)^{-\frac{2\epsilon}{\zeta}},
  \hspace{37mm} {\rm for}\hspace{7mm} n=-2.
 \end{array}\right.
\end{equation}

Substituting $\phi^{\frac{n}{2}}\dot{\phi}$ from \eqref{AES-1} into
equations \eqref{A-1}, \eqref{rho-sol1} and \eqref{H-sol1}, we can easily
obtain the potential, energy density and the scale factor in terms of the cosmic time:
\begin{eqnarray}
\label{Ex1-Vt}
V(t)&=&V_i\left(\frac{t}{t_i}\right)^{-2},\hspace{18mm}
 V_i\equiv \left[\frac{4(2-{\cal W}\Gamma)}{\zeta^2\Gamma}\right]\frac{1}{t_i^{2}},\\\nonumber\\
\label{Ex1-ro-phi}
\rho_\phi(t)&=&\left[\frac{8}{(D-2)\Gamma \zeta^2}\right]\,\frac{1}{t^2},\\\nonumber\\
\label{AES-3}
a(t)&=&a_i\left(\frac{t}{t_i}\right)^{\beta}, \hspace{18mm} \beta\equiv\frac{2\epsilon}{(D-1){\cal W}\Gamma},
\end{eqnarray}
which are valid for all values of $n$. In \eqref{AES-3}, $a_i$ is an integration constant. For $\epsilon=1$ and $\beta>1$ one can obtain a power-law inflation \cite{LM85, RSM22, GC07} at early times.
It should be noted that, because $\beta$ is a constant in this scenario, assuming that it takes values greater than one, the universe accelerates indefinitely. The number of e-folds, $\mathcal{N}$, during the period of inflation is:
\begin{equation}
\label{e-fol}
\mathcal{N}=\ln\left[{\frac{a(t_{\textit{f}})}{a(t_{\textit{i}})}}\right]=\int^{t_{\textit{f}}}_{t_i}dt H(t) \sim\frac{2}{(D-1){\cal W}\Gamma}\ln{t_{\textit{f}}},
\end{equation}
which depends on the number of the dimensions of the universe and ${\cal W}$.
Such a solution has been also obtained using other gravitational frameworks \cite{RFM14,RS21,RCMJ22}.

Employing relations \eqref{AES-2} and \eqref{Ex1-Vt},
we can obtain the potential in terms of the scalar field:
\begin{equation}\label{Ex1-Vphi}
V(\phi)=\left \{
 \begin{array}{c}
V_i
{\exp}\left[\left(\frac{2\epsilon\zeta}{n+2}\right)
\left(\phi^{\frac{n+2}{2}}-\phi_i^{\frac{n+2}{2}}\right)\right],
 \hspace{12mm} {\rm for}\hspace{10mm} n\neq-2,\\\\
 V_i \left(\frac{\phi}{\phi_i}\right)^{\epsilon\zeta},
  \hspace{38mm} {\rm for}\hspace{10mm} n=-2.
 \end{array}\right.
\end{equation}
Using relations  \eqref{AES-1} and \eqref{Ex1-Vt}, we showed that
\begin{eqnarray}
\label{Ex1-p-phi}
p_\phi(t)=\left[\frac{8({\cal W}\Gamma-1)}{\zeta^2\Gamma}\right]\,\frac{1}{t^2},\hspace{13mm} \forall n.
\end{eqnarray}

In the next section we will analyze the behavior of some of the physical quantities obtained above and compare them with their corresponding NC counterparts.

\section{Noncommutative classical SB cosmology}
\label{SecIV}
In this section we assume that the NC parameter is
nonzero and takes very small values; clearly, when
it vanishes, all field equations are reduced to those of the commutative case.

To specify the behavior of physical quantities for the classical NC model, we can consider two methods as follows.

\subsection{First method: numerical investigation}
\label{First method}
In this method, we first solve the equation \eqref{H-sol1} to relate the scale factor to the scalar field. If we then plug the resulting relation into \eqref{AE-1}, we get a differential equation for just $\phi(t)$. It turns out that the solution of such a complicated equation is not analytically feasible. Therefore we have to use a numerical method to analyze the behavior of the required quantities.
Because we will be studying the dynamics of the problem below, we will refrain from illustrating the behavior of all physical quantities and will only briefly describe the results here.

Our numerical investigations show that we can reach an inflationary epoch at earlier times, such that after a short time the scale factor transitions gracefully from an acceleration phase and enters to a deceleration phase, which is due to a radiation--dominated universe. In addition, the nominal condition is relevant for the causal physics of inflation is met. Other more appropriate methods, which are discussed below, can be utilised to demonstrate the above results.

\begin{figure}
\centering\includegraphics[width=2.5in]{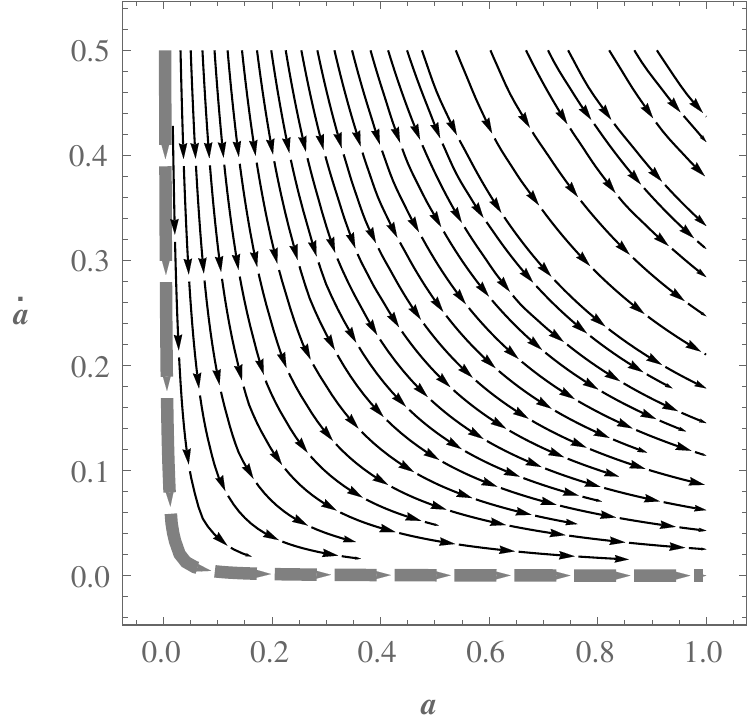}
\hspace{5mm}
\centering\includegraphics[width=2.5in]{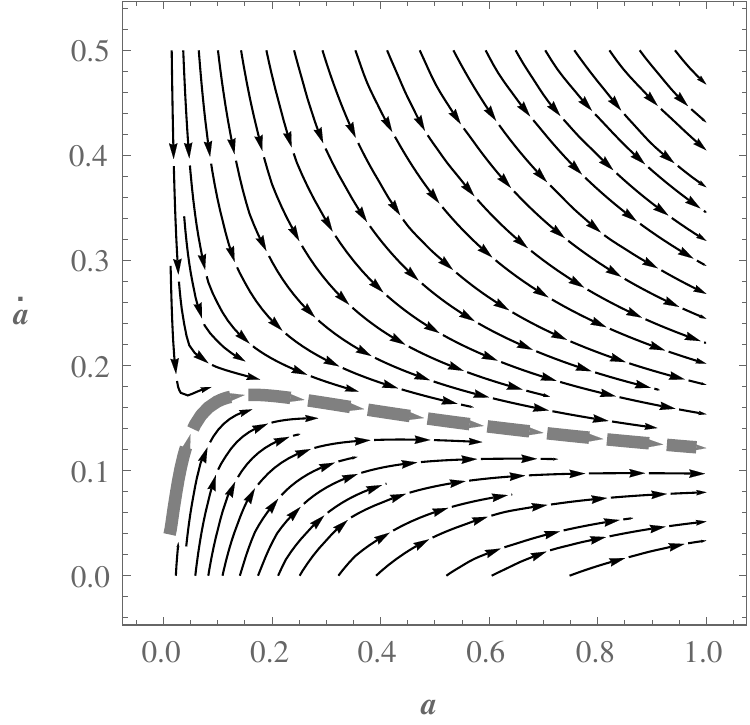}
\caption{Phase space portrait associated with the commutative
case (left panel) and NC case (the right panel assuming $\theta=-0.0001$), respectively, according to equation (\ref{y-dyn-arb-n}).
We have assumed $n=3$, ${\cal W}=2$, $\Gamma=0.8$, $8\pi G=1$, $a_i=0.005$, $\phi_i=1$ and $D=4$.
}
\label{PhaseP-n}
\end{figure}

\begin{figure}
\centering\includegraphics[width=2.5in]{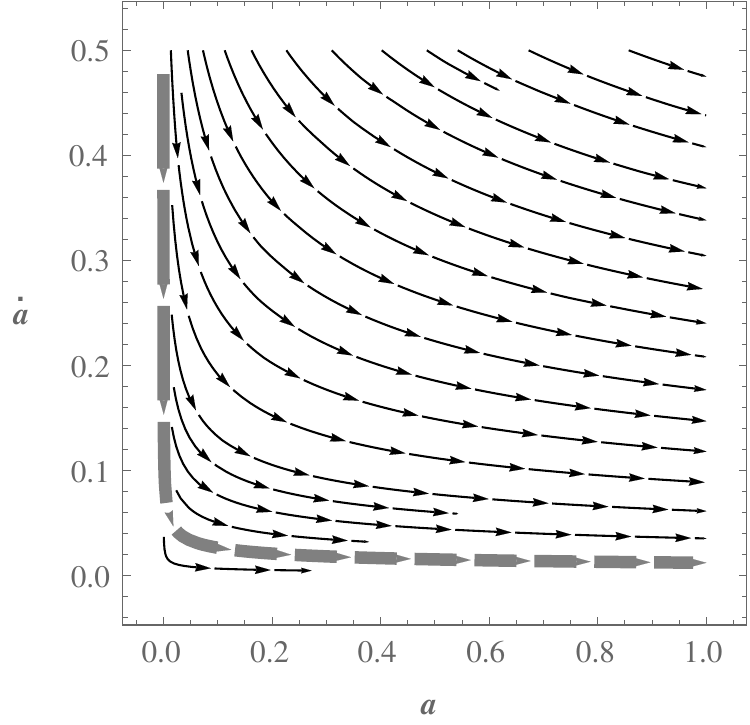}
\hspace{5mm}
\centering\includegraphics[width=2.5in]{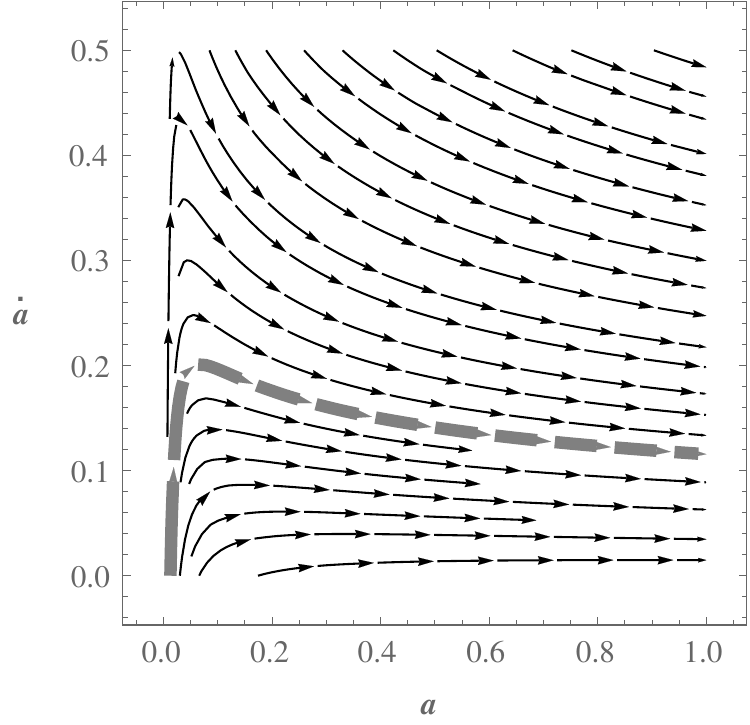}
\caption{Phase space portrait associated with the commutative
case (left panel) and NC case (the right panel assuming $\theta=-0.01$) for the case $n=-2$, according to equation \eqref{y-dyn-2}.
We have assumed ${\cal W}=0.09$, $\Gamma=9.9$, $8\pi G=1$, and $D=4$.
 }
\label{PhaseP-2}
\end{figure}

To prove that our model is a successful inflationary model, it is easy to provide its corresponding dynamical system as
\begin{equation}
\dfrac{d\tilde{a}}{da}=\left[1-\frac{\mathcal{W}}{(D-2)h_0^2}\right]\left(\frac{\tilde{a}}{a}\right)-\frac{\theta a^{2-D}\left[\phi_i^{\frac{n+2}{2}}+\frac{n+2}{2\alpha}\ln\left(\frac{a}{a_i}\right)\right]^{\frac{D+2}{D-2}}}{2(D-1)(D-2)h_0}
,\hspace{3mm}n\neq-2,
\label{y-dyn-arb-n}
\end{equation}
and
\begin{equation}
\dfrac{d\tilde{a}}{da}=\left[1-\frac{\mathcal{W}}{(D-2)h_0^2}\right]\left(\frac{\tilde{a}}{a}\right)-\frac{\epsilon\theta a^{2-D}}{2(D-1)(D-2)h_0},\hspace{17mm}n\neq-2,
\label{y-dyn-2}
\end{equation}
where $\tilde{a}=\dot{a}$, and we have used \eqref{asli2} and \eqref{H-sol1}.
Our numerical endeavors have shown that the model is very sensitive to the values of the parameters and the constants of integration ($a_i$ and $\phi_i$) and initial conditions (ICs). However, we always have an inflationary epoch that corresponds to certain sets of reasonable values. Let us focus more precisely on the phase portrait of equations \eqref{y-dyn-arb-n} and \eqref{y-dyn-2}, by which we compare the commutative model and the corresponding NC model using the same values for the common parameters, integration constants and ICs. As the left panels of figures \ref{PhaseP-n} and \ref{PhaseP-2} show, for the commutative case, $\dot{a}$ always decreases as scale factor increases, indicating a decelerating universe. For an NC model fitted to very small values of the NC parameter, we get an interesting behavior for the scale factor. Specifically, as shown in Figures
\ref{PhaseP-n} and \ref{PhaseP-2}
(see right panels), we get an inflationary epoch in very early times. Thereafter, there is an suitable mechanism for exiting from the accelerating phase gracefully and thus a transition into the deceleration phase.
It is worth noting that the application of the exponential potential is useful in most classical models to obtain exact solutions for an inflationary epoch. However, these models, without dictating some additional conditions, usually do not have a suitable mechanism to end the inflationary phase. We emphasize that the problem has been solved in our NC model.

It is important to note that the mentioned statements are also valid for $n=-2$, see figure \ref{PhaseP-2}.

Since we cannot reconstruct the Lagrangian belonging to our NC model, let us compare our model with the well-known Starobinsky model \cite{S80} (see also \cite{V85}) at the level of the associated field equations.
Employing equations \eqref{asli2} and \eqref{H-sol1}, it is easy to show that, the evolution of the scale factor for our NC model \textit{for $n\neq-2$} is
\begin{eqnarray}\nonumber
\frac{\dddot{a}}{a}&+&\left[\frac{2{\cal W}-3(D-2)h_0^2}{(D-2)h_0^2}\right]\left(\frac{\dot{a}\ddot{a}}{a^2}\right)
-2\left[\frac{{\cal W}-(D-2)h_0^2}{(D-2)h_0^2}\right]\left(\frac{\dot{a}}{a}\right)^3\\\nonumber\\\nonumber
&+&
\frac{\theta a^{2-D}}{2(D-1)(D-2)h_0}\left[\phi_i^{\frac{n+2}{2}}+\frac{n+2}{2\alpha}\ln\left(\frac{a}{a_i}\right)\right]^{\frac{D+2}{D-2}}
\Bigg\{\frac{\ddot{a}}{a}+\left(\frac{\dot{a}}{a}\right)^2
\\\nonumber\\
&\times&\left(\frac{(D+2)(n+2)}{2(D-2)h_0}\left[\phi_i^{\frac{n+2}{2}}+\frac{n+2}{2\alpha}\ln\left(\frac{a}{a_i}\right)\right]^{-1}-(D-1)\right)\Bigg\}=0
\label{evol-a-2}
\end{eqnarray}
and \textit{for $n=-2$} is
\begin{eqnarray}\nonumber
\frac{\dddot{a}}{a}&+&\left[\frac{2{\cal W}-3(D-2)h_0^2}{(D-2)h_0^2}\right]\left(\frac{\dot{a}\ddot{a}}{a^2}\right)
-2\left[\frac{{\cal W}-(D-2)h_0^2}{(D-2)h_0^2}\right]\left(\frac{\dot{a}}{a}\right)^3\\\nonumber\\
&+&
\frac{\epsilon\theta a^{2-D}}{2(D-1)(D-2)h_0}\left[\frac{\ddot{a}}{a}+(1-D)\left(\frac{\dot{a}}{a}\right)^2\right]=0.
\label{evol-a-3}
\end{eqnarray}
For the special case that our model is reduced to the Einstein scalar field system with canonical kinetic term, the equations \eqref{evol-a-2} and \eqref{evol-a-3} can correspond to the evolution equation for the scale factor, which was obtained in Ref. \cite{V85} and justify the presence of the NC parameter and the integration constants (of our model) by the presence of the coefficients introduced in \cite{V85}.

\subsection{Second method}
\label{Second method}

In this subsection, we would like to explore another interesting approximate solution for our NC model.
Let us examine the intriguing NC inflationary model \cite{R11}:
\begin{eqnarray}
\label{inf-1}
a(t)=a_i e^{\sqrt{\frac{\pi  \Theta }{2}} H_i \text{erf}\left(\frac{t}{\sqrt{2 \Theta }}\right)},
\end{eqnarray}
where $H_i$ is the value of the Hubble parameter at $t=t_i$, $\Theta\propto \theta^2$, $a_i$ is a constant and
\begin{eqnarray}
\label{erf}
 \text{erf}(x)=\frac{2}{\sqrt{\pi}}\int_0^{x}\,e^{-\tilde{x}^2}d\tilde{x}\ .
\end{eqnarray}
Substituting $a(t)$ from \eqref{inf-1} into \eqref{H-sol1}, we obtain
\begin{eqnarray}
\label{inf-2}
\phi(t)=\left[\frac{n+2}{4} \left(C\pm   \sqrt{2\pi  \Theta }\,  \text{erf}\left(\frac{t}{\sqrt{2 \Theta }}\right)\right)\right]^{\frac{2}{n+2}},
\end{eqnarray}
where $n\neq-2$ and $C$ is a constant of integration.

\begin{figure}
\centering\includegraphics[width=2.5in]{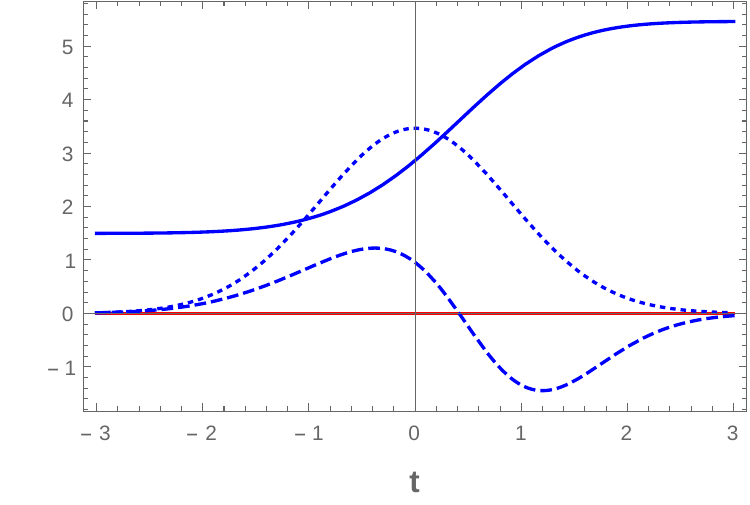}
\hspace{5mm}
\centering\includegraphics[width=2.5in]{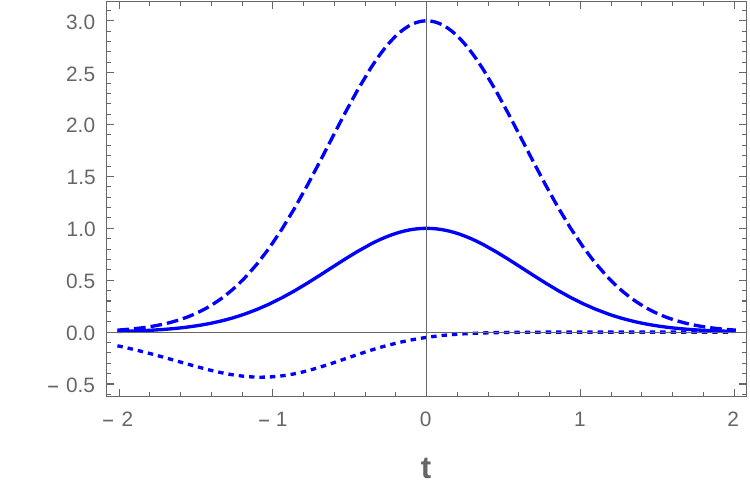}
\caption{Left panel: The time behavior of the scale factor $a$ (solid curve),  $\ddot{a}$ (dashed curve) and the Hubble parameter $H$ (dotted curve). Right panel: The qualitative behavior of the $\rho_\phi$ (solid curve), $p_\phi$ (dashed curve), and $p_{_{\rm nc}}$ (dotted curve). Some quantities have been re-scaled for clarity. Moreover, we have assumed ${\cal W}=4$, $a_i=2.86$, $C=2$, $n=-1$, $\Theta=0.8$, $\Gamma=1$, $8\pi G=1$, and $D=4$.
 }
\label{NC-infl}
\end{figure}

\begin{figure}
\centering\includegraphics[width=2.5in]{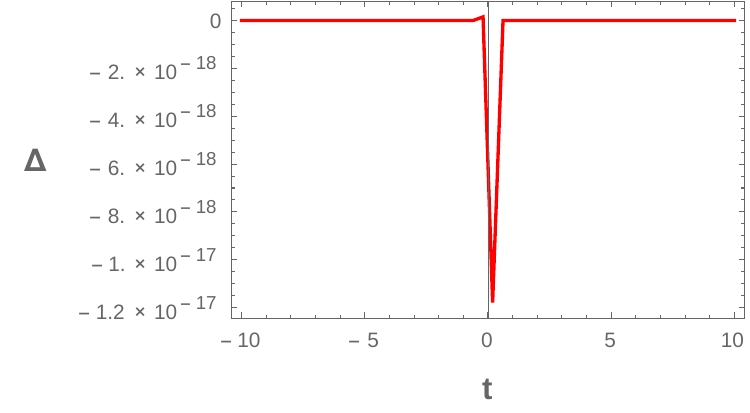}
\caption{The time behavior of the quantity $\Delta$ (see equation \eqref{cons-eq}) for $\Theta=0.0008$. We have taken the values of the other parameters as used in figure \ref{NC-infl}.
 }
\label{Del}
\end{figure}

For such a solution, as an example, in figure \ref{NC-infl}, we have plotted the qualitative behavior of some relevant physical quantities in terms of time varying from $-\infty$ to $+\infty$. As seen, the Hubble parameter reaches its maximum at $t=0$, while $\ddot{a}$ changes sign and turn to be negative when the comoving Hubble length $(a H)^{-1}$ and the effective (NC) pressure $p_{_{\rm nc}}$ reach their minimum.
Our numerical results reveal that the smaller $\Theta$, the better equation \eqref{cons-eq} (or \eqref{AE-1}) is satisfied, see for instance figure \ref{Del}.

\section{Commutative quantum cosmology}
\label{SecV}
In this and the following sections, all variables will be treated as operators, so those with the index $c$ will be associated to the commutative quantum model, while those without an index will be related to the NC one.
For the sake of simplicity, however, we omit the hats on the operators.

First, let us investigate the commutative quantum SB-FLRW cosmological model in $D$ dimensions. To get the commutative WDW equation, we should consider the canonical quantization of the classical super--Hamilton constraint \eqref{constraint-asli} based on the ordinary Heisenberg-Weyl algebra:
\begin{eqnarray}
\label{c-commutators}
\left[{a}_c, {\phi}_c\right]=0=\left[ \Pi_{a_c}, \Pi_{\phi_c}\right],
\nonumber\\\nonumber\\
\label{c-commutators}
\left[{a}_c,  \Pi_{a_c}\right]=i=\left[{\phi}_c,  \Pi_{\phi_c}\right]~,
\end{eqnarray}
where $\Pi_{a_c}$ and $\Pi_{\phi_c}$ are the fundamental
momentum operators conjugate to ${a_c}$ and ${\phi_c}$, respectively.

The canonical quantization of the
Hamiltonian constraint \eqref{constraint-asli} gives the WDW equation for the wave
function of the universe associated with the commutative case as
\begin{eqnarray}\nonumber
 &\left[-\Pi_{a_c}^2
+(D-1)(D-2)\mathcal{W}^{-1}a_c^{-2}\phi_c^{-n} \Pi_{\phi_c}^2\right]\psi(a_c,\phi_c)\\\nonumber\\
&+\left[4(D-1)(D-2)a_c^{2(D-2)}V(\phi_c)\right]\psi(a_c,\phi_c)=0,
\label{constraint}
\end{eqnarray}
where we
assumed there is no ordinary matter.

By substituting $\Pi_{a_c}$ and $\Pi_{\phi_c}$ from
\begin{equation}
\label{Fac-ord}
\Pi_{a_c}^2\rightarrow -a_c^{-p}\frac{\partial }{\partial a_c}
\left(a_c^p\frac{\partial }{\partial a_c}\right), \hspace{8mm}
\Pi_{\phi_c}^2 \rightarrow -\frac{{\partial }^2}{\partial \phi_c ^2},
\end{equation}
into \eqref{constraint} we get
\begin{multline}
\label{wdw2}
\left[\frac{\partial^2}{\partial a_c^2}+
\frac{p}{a_c}\frac{\partial}{\partial a_c}-
\frac{(D-1)(D-2)}{\mathcal{W} a_c^2}\phi_c^{-n}\frac{\partial^2}{\partial \phi_c^2}\right]\psi(a_c,\phi_c)\\
+\left[4(D-1)(D-2)\mathcal{V}_0^2a_c^{2D-4}V(\phi_c)\right]\psi(a_c,\phi_c)=0,
\end{multline}
where $p$ is an ordering parameter that is included for the factor-ordering ambiguity \cite{V87}.
In Ref. \cite{CM00}, assuming $p=-1$, a quantum creation of universes
with compact spacelike sections corresponding to a particular case of the model \eqref{wdw2},
where $D=4$, $n=0$ and $\mathcal{W}=1/2$ has been studied in detail.


Throughout this work we will apply the simplest factor ordering of the operators, i.e. we set $p=0$.


In this section, let us proceed our investigation with an exponential scalar potential.
 For simplicity, we use a suitable transformation as
\begin{eqnarray}\label{a-alpha}
{a_c}=e^\alpha,
\end{eqnarray}
(where $\alpha=\alpha_c$ is a commutative variable) hence the corresponding conjugate momenta are related as
\begin{eqnarray}\label{pa-alpha}
\Pi_{a_c}=e^{-\alpha}\Pi_{\alpha}.
\end{eqnarray}
Therefore, the WDW equation is rewritten as
\begin{eqnarray}\nonumber
&-&\left[\frac{\partial^2}{\partial \alpha^2}
-\frac{(D-1)(D-2)}{\mathcal{W}}\phi_c^{-n}\frac{\partial^2}{\partial \phi_c^2}\right]\psi(\alpha,\phi_c)\\\nonumber\\
&+&4(D-1)(D-2)\exp[2(D-1)\alpha]V(\phi_c)\psi(\alpha,\phi_c)=0~.
\label{C-WDW-1}
\end{eqnarray}

 Let us restrict ourselves to the case $n\neq-2$ and use the same exponential
 potential obtained in Section \ref{SecIII}, i.e., equation \eqref{Ex1-Vphi}, where for simplicity we set $\phi_i=0$. Using another transformation
 \begin{eqnarray}
 \phi_c\equiv\left[\frac{(D-1)(D-2)}{\mathcal{W}}\right]^{\frac{1}{n+2}}\varphi,
\label{Phi-phi}
\end{eqnarray}
(where $\varphi=\varphi_c$ is a commutative variable) the WDW equation is written as
\begin{eqnarray}
\left[-\frac{\partial^2}{\partial \alpha^2}
+\varphi^{-n}\frac{\partial^2}{\partial \varphi^2}
+4(D-1)(D-2) e^{2(D-1)\alpha}V(\varphi)\right]\psi(\alpha,\varphi)=0~,
\label{C-WDW-2}
\end{eqnarray}
 where
 \begin{eqnarray}
V(\varphi)=V_i \exp[-\sigma \varphi^{\frac{n+2}{2}}], \hspace{10mm} \sigma\equiv \frac{\epsilon (D-1)\sqrt{8\mathcal{W} \Gamma }}{n+2}.
\label{C-pot}
\end{eqnarray}
Employing transformations
 \begin{eqnarray}
x=-2(D-1)\alpha+\sigma \varphi^{\frac{n+2}{2}}, \hspace{10mm}   y=-\alpha+\frac{8(D-1)}{\sigma (n+2)^2} \varphi^{\frac{n+2}{2}},
\label{x-y}
\end{eqnarray}
it is easy to show that equation
\eqref{C-WDW-1} reduces to
\begin{eqnarray}
\left[\frac{\partial^2}{\partial x^2}
-\frac{\partial^2}{\partial\tilde{y}^2}
-\upsilon e^{-x}\right]\psi(x,\tilde{y})=0~,
\label{C-WDW-3}
\end{eqnarray}
where
\begin{eqnarray}
\tilde{y}\equiv\left[{\frac{(n+2)\sigma}{2}}\right]y, \hspace{10mm}
\upsilon\equiv-\frac{4(D-1)(D-2)V_i}{\left(\frac{(n+2)\sigma}{2}\right)^2-4(D-1)^2}.
\label{def-ytild}
\end{eqnarray}
Assuming $\psi=X(x)Y(\tilde{y})$, equation \eqref{C-WDW-3} yields
\begin{eqnarray}
\frac{1}{X}\frac{d^2X}{dx^2}-\upsilon e^{-x}=\frac{1}{Y}\frac{d^2Y}{d\tilde{y}^2}
\label{C-WDW-4}
\end{eqnarray}
The solutions for these differential equations are:
\begin{equation}\label{C-X}
X(x)=\begin{cases}
J_{i\eta}\left(\pm2\sqrt{\upsilon} e^{-\frac{x}{2}}\right)+J_{-i\eta}\left(\pm2\sqrt{\upsilon} e^{-\frac{x}{2}}\right),
 \hspace{5mm} {\rm for}\hspace{5mm} |(n+2)\sigma|<2(D-1),\\\\
I_{i\eta}\left(\pm2\sqrt{\upsilon} e^{-\frac{x}{2}}\right)+K_{-i\eta}\left(\pm2\sqrt{\upsilon} e^{-\frac{x}{2}}\right), \hspace{5mm} {\rm Otherwise},
 \end{cases}
\end{equation}
\begin{equation}\label{C-Y}
Y(\tilde{y})=C_1 \cos\left({\frac{\eta\tilde{y}}{2}}\right)+C_2\sin\left({\frac{\eta\tilde{y}}{2}}\right),
\end{equation}
where $C_1$ and $C_2$ are integration constants, $\eta$
is a separation constant, $I_{i\eta}$ and $K_{i\eta}$ are the modified Bessel Functions.
In order to obtain a normalizable wave function we should establish wave packets to construct a Gaussian state.

In order to obtain the classical limit of the wavefunction, let us use the WKB--like method. Let us consider
\begin{equation}\label{WKB-1}
\psi(x,\tilde{y})=\exp{\left[-S(x,\tilde{y})\right]}
\end{equation}
and use the standard conditions
\begin{eqnarray}\label{WKB-2}
\left(\frac{\partial S}{\partial x}\right)^2>>\Big|\frac{\partial^2 S}{\partial x^2}\Big|,
\hspace{10mm}\left(\frac{\partial S}{\partial \tilde{y}}\right)^2>>\Big|\frac{\partial^2 S}{\partial \tilde{y}^2}\Big|.
\end{eqnarray}
Therefore, equation \eqref{C-WDW-3} reduces to
\begin{eqnarray}\label{WKB-3}
\left(\frac{\partial S}{\partial x}\right)^2-\left(\frac{\partial S}{\partial \tilde{y}}\right)^2-\upsilon e^{-x}=0.
\end{eqnarray}
Assuming $S=S_x(x)S_{\tilde{y}}(\tilde{y})$, the solutions of \eqref{WKB-3} are:
\begin{eqnarray}\label{WKB-4}
S_{\tilde{y}}=s={\rm constant}, \hspace{10mm} S_x=\pm\frac{2}{s\sqrt{\upsilon}}e^{-\frac{x}{2}}
\end{eqnarray}
and therefore
\begin{eqnarray}\label{WKB-5}
\Pi_{\tilde{y}}\equiv\frac{\partial S}{\partial \tilde{y}}=0, \hspace{10mm}\Pi_x\equiv\frac{\partial S}{\partial x}=\pm\frac{1}{\sqrt{\upsilon}}e^{-\frac{x}{2}}.
\end{eqnarray}

Reconsidering the variables $\alpha$ and $\varphi$, we can easily obtain
\begin{eqnarray}\label{p-alfa}
\Pi_\alpha=\pm\frac{2(D-1)}{\sqrt{\upsilon}}\exp{\left[(D-1)\alpha-\frac{\sigma}{2}\varphi^{\frac{n+2}{2}}\right]},\\\nonumber\\
\label{p-phi}
\Pi_\varphi=\mp\left[\frac{\sigma(n+2)}{2}\right]\varphi^{\frac{n}{2}}\exp{\left[(D-1)\alpha-\frac{\sigma}{2}\varphi^{\frac{n+2}{2}}\right]}.
\end{eqnarray}
Finally, using equations \eqref{momenta}, \eqref{a-alpha}, \eqref{pa-alpha} and \eqref{Phi-phi}, after some manipulation, we get back the same results for scale factor and scalar field obtained in section \ref{SecIII}, see equations \eqref{AES-2} and \eqref{AES-3} for the case $n\neq-2$ with $\phi_i=0$.


\section{Noncommutative quantum cosmology}
\label{SecVI}
As mentioned, solving the NC-WDW equation associated with the
model investigated in Section \ref{SecII} is a very complicated procedure.
Therefore, from the beginning, let us restrict our
attention to a NC quantum model corresponding to $\{\Pi_a,\Pi_{\phi }\}=\theta$, $n=0$ and $V=V_0={\rm constant}$.
Therefore, the equations \eqref{diff.eq1} and \eqref{diff.eq4} yield a constant of motion:
\begin{eqnarray}\label{con-of-motion}
{\Pi}_{\phi}+\theta a=C,
\end{eqnarray}
which will assist us to proceed our discussions in this Section.

Performing the quantization of the classical algebra \eqref{deformed}
for this particular case yields a NC Heisenburg-Weyl algebra as
\begin{eqnarray}   \label{NC-WDW}
\left[{a},  \Pi_{a}\right]=i, \hspace{0.2 cm} \left[{\phi}, \Pi_{\phi}\right]=i\hspace{0.2 cm},
\hspace{0.2 cm}\left[\Pi_{a}, \Pi_{\phi}\right]=i\theta,
 \end{eqnarray}
 where the other commutators vanish. We should remind that the
 operators without index c are those associated with the NC quantum model.

The best way to represent the algebra \eqref{NC-WDW} is to use
the transformations
 \begin{eqnarray}\label{D-map-1}
{a}& =& {a}_{c}, \hspace{35mm} {\phi} = {\phi}_{c}~,\\\nonumber\\
\label{D-map}
 \Pi_{a}&=& \Pi_{a_c} + \frac{\theta}{2} {\phi}_{c}, \hspace{20mm}
 \Pi_{\phi}= \Pi_{\phi_c}- \frac{\theta}{2} {a}_{c}~,
\end{eqnarray}
which restores the standard Heisenberg algebra. Employing
transformations \eqref{D-map-1} and \eqref{D-map} we may
consider \eqref{NC-WDW} as an algebra of operators that act on the usual Hilbert space $L^2(\mathbb{R}^2)$.

Using the transformations \eqref{D-map} provides a representation that we can write the NC-WDW equation as
\begin{eqnarray}\nonumber
\left[-\left(-i \frac{\partial}{\partial {a_c}}+\frac{\theta}{2}\phi_{c}\right)^2
+\frac{(D-1)(D-2)}{\mathcal{W}}a_c^{-2}\left(-i  \frac{\partial}{\partial {\phi_c}}-\frac{\theta}{2}a_c\right)^2\right]\psi(a_c,\phi_c)\\
\label{NC-WDW-2}
+4(D-1)(D-2)V_0a_c^{2(D-2)}\psi(a_c,\phi_c)=0~,
\end{eqnarray}
which explicitly depends on the NC parameter.
To solve this complex equation, we invoke the NC quantum version of the constant of motion \eqref{con-of-motion},
which can be also written as
\begin{equation}
\label{Q-cons-2}
{C}={\Pi}_{\phi_c}+\frac{\theta}{2}{a}_c,
\end{equation}
where we used the transformations \eqref{D-map}.
It is straightforward to show that ${C}$ commutes with NC Hamiltonian (in the constraint space of states):
\begin{equation}\label{Q-cons-3}
[{\Pi_{\phi}}+\theta{a},H]=0.
\end{equation}
Namely, any solution of equation \eqref{NC-WDW-2} is also a simultaneous
eigenstate of $C$. Let us consider $\psi_m(a_c,\phi_c)$ as an eigenstate of the operator ${C}$ with eigenvalue $m\in \mathbb{R}$:
\begin{eqnarray}
\label{Psi}
\left(-i\frac{\partial}{\partial\phi_c}+\frac{\theta}
{2}a_c\right)\psi_m(a_c,\phi_c)=m \psi_m(a_c,\phi_c)~,
\end{eqnarray}
where we used \eqref{Q-cons-2}.

It is easy to show that a solution for \eqref{Psi} is:
\begin{equation}
    \label{sol.cons}
\psi_m(a_c,\phi_c)=A(a_c)\exp{\left[i\left(m-\frac{\theta}{2}a_c\right)\phi_c\right]}~.
\end{equation}

Substituting $\psi_m(a_c,\phi_c)$ from \eqref{sol.cons} into \eqref{NC-WDW-2} yields a second order ordinary differential equation:
\begin{equation}
\label{NC-WDW-3}
A''(a_c)+(D-1)(D-2)\left[\mathcal{W}^{-1}a_c^{-2}\left(m-\theta a_c\right)^2+4V_0a^{2(D-2)}\right]A(a_c)=0~,
\end{equation}
 where a prime denotes a differentiation with respect to its argument.
 It is seen that equation \eqref{NC-WDW-3}, in turn, depends on the NC parameter $\theta$ and the eigenvalue $m$.
 \begin{figure}
\centering\includegraphics[width=4.5in]{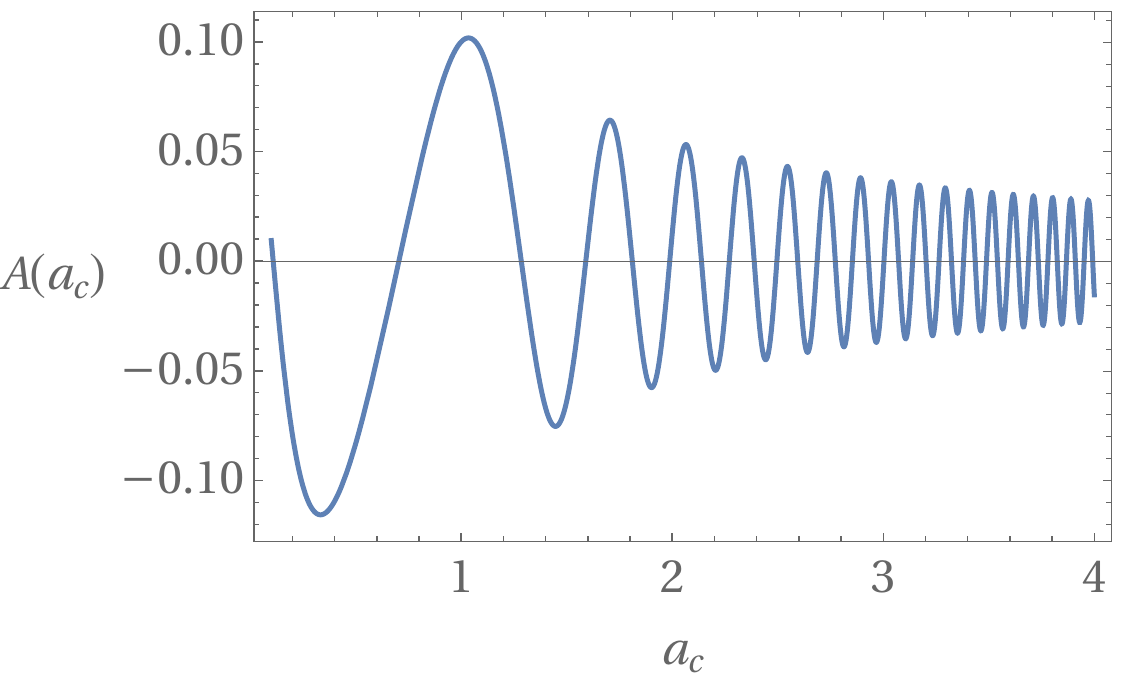}
\caption{Numerical solution of Eq. \ref{NC-WDW-3}, with initial conditions $A(1)=0.1$, $A'(1)=0.1$.
Moreover, we have assumed $m=1$, $\theta=-0.01$, $V_0=1$, ${\cal W}=2$  and $D=4$.
}
\label{NC-QC-num}
\end{figure}

The equation \eqref{NC-WDW-3} cannot be solved analytically except
under certain circumstances. Let us first investigate such particular analytical solutions.
In the limit of very small $a_c$, keeping the terms of order up to $\mathcal{O}(a_c^{-2})$, Eq. \eqref{NC-WDW-3} reduces to
 \begin{eqnarray}
  \label{NC-WDW-4}
 A''(a_c)+\left[(D-1)(D-2)\mathcal{W}^{-1}a_c^{-2} m^2\right]A(a_c)&=&0~, \hspace{5mm} {\rm for } \hspace{5mm} m\neq0,\\\nonumber\\
 \label{NC-WDW-5}
 A''(a_c)+\left[(D-1)(D-2)\mathcal{W}^{-1} \theta^2\right]A(a_c)&=&0~, \hspace{5mm} {\rm for } \hspace{5mm} m=0.
 \end{eqnarray}
For the latter, we can see that the full wave function $\psi_m(a_c,\phi_c)$ will carry effects from the NC parameter.

\begin{figure}
\centering\includegraphics[width=2.5in]{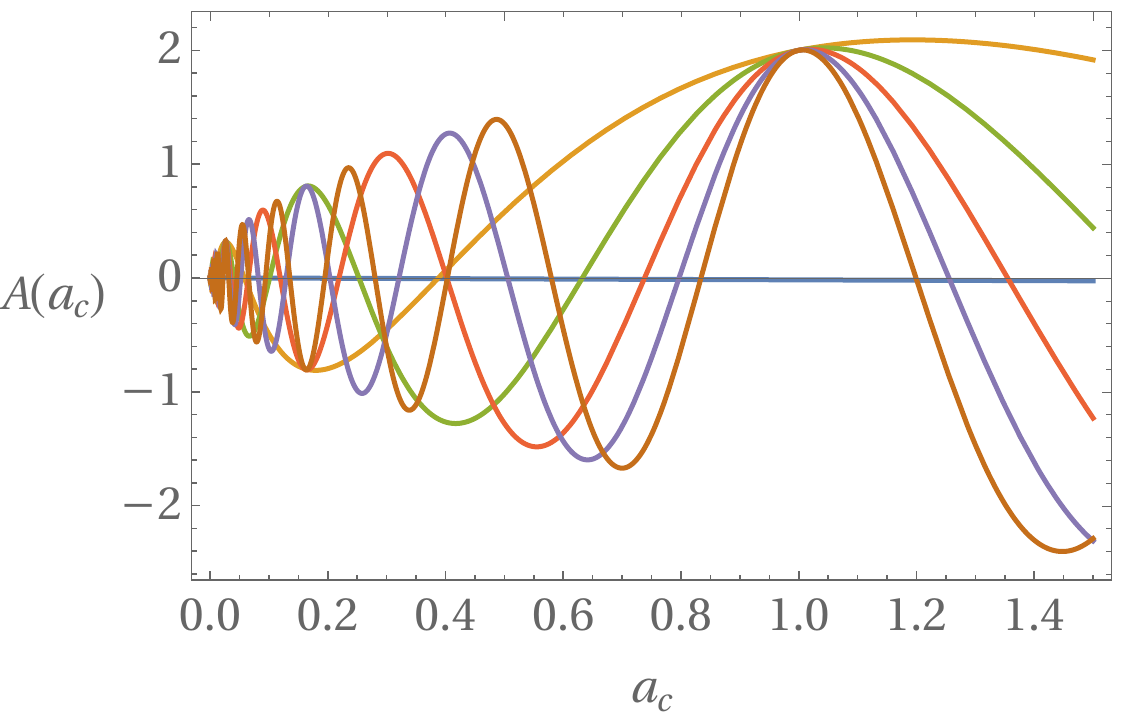}
\hspace{0.1cm}
\centering\includegraphics[width=2.5in]{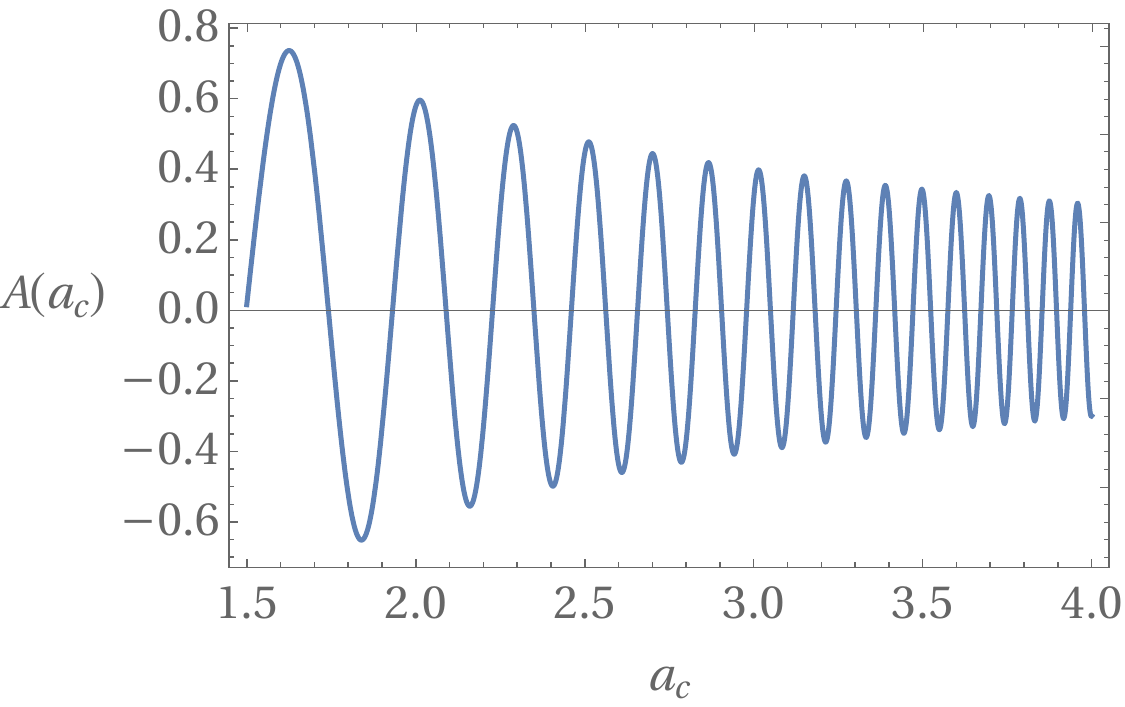}
\caption{In the left panel we represent analytical solutions of
Eqs. \eqref{ansol1}-\eqref{ansol2} for the eigenvalue $m=0,1,\ldots,5$.
In the right panel we depict the behaviour of the analytical solution \ref{ansol3}.
Moreover, we have also assumed $\theta=-0.01$, $V_0=1$, ${\cal W}=2$  and $D=4$.
}
\label{NC-QC-ana1}
\end{figure}

 The analytical solution for Eq. \eqref{NC-WDW-4} is:
 \begin{equation}
 \label{ansol1}
A(a_c)=
a_c^{\frac{1}{2}-\frac{m \sqrt{(D-1)(D-2)} \sqrt{\frac{\mathcal{W}}{m^2 (D-1)(D-2)}-4}}{2 \sqrt{\mathcal{W}}}} \left(c_2 \: a_c^{\frac{m \sqrt{(D-1)(D-2)} \sqrt{\frac{\mathcal{W}}{m^2 (D-1)(D-2)}-4}}{\sqrt{\mathcal{W}}}}+c_1\right) ,
 \end{equation}
 where $c_1$ and $c_2$ are integration constants, and can be adjusted to make $\psi_m(a_c,\phi_c) \rightarrow 0$ for $a_c \rightarrow 0$.
 Moreover, the analytical solution for Eq. \eqref{NC-WDW-5} is given by
 \begin{equation}
  \label{ansol2}
A(a_c)= c_2 \sin \left(\sqrt{3} \theta  a_c \right)+c_1 \cos \left(\sqrt{3} \theta  a_c \right).
 \end{equation}
 In the left panel of Figure \ref{NC-QC-ana1}, we depict some
 particular solutions for Eqs. \eqref{ansol1} and \eqref{ansol2} to
 cover the values for the eigenvalue $m=0,1,\ldots,5$.
 We adjusted the integration constants $c_1$  and $c_2$ such that the
 wave function starts at zero for $a_c=0$. We have, initially, an
 oscillatory behaviour that gets amplified as $a_c$ grows for all the modes (different $m$).

 Furthermore, for very large $a_c$, by keeping only the term of order $\mathcal{ O}(2(D-2))$, Eq. \ref{NC-WDW-3} reduces to
 \begin{equation}
  \label{NC-WDW-6}
 A''(a_c)+\left[4(D-1)(D-2)V_0a^{2(D-2)}\right]A(a_c)=0~.
 \end{equation}
 An analytical solution for \eqref{NC-WDW-6} is:
 \begin{equation}
 \label{ansol3}
 \begin{split}
 A(a_c)= \frac{\sqrt[12]{(D-1)(D-2)}  \sqrt{a_c}}{\sqrt[6]{3}}
 \Biggl[ & c_1 \Gamma \left(\frac{5}{6}\right) J_{-\frac{1}{6}}\left(\frac{2}{3} \sqrt{(D-1)(D-2)}  a_c^3\right) \\
 + & c_2 \Gamma \left(\frac{7}{6}\right) J_{\frac{1}{6}}\left(\frac{2}{3} \sqrt{(D-1)(D-2)}  a_c^3\right)\Biggr].
 \end{split}
 \end{equation}
 We observe that this solution does not depend on either the NC parameter $\theta$ or
 the eigenvalue $m$ for large values of $a_c$. Moreover, according to the right
 panel of Figure \ref{NC-QC-ana1}, where a solution of Eq. \ref{ansol3} was depicted,
 the behavior of the analytical solutions for large $a_c$ do match with the
 behaviour found for the numerical solution of Eq. \ref{NC-WDW-3} shown in Figure \ref{NC-QC-num}.



\begin{figure}
\centering\includegraphics[width=2.5in]{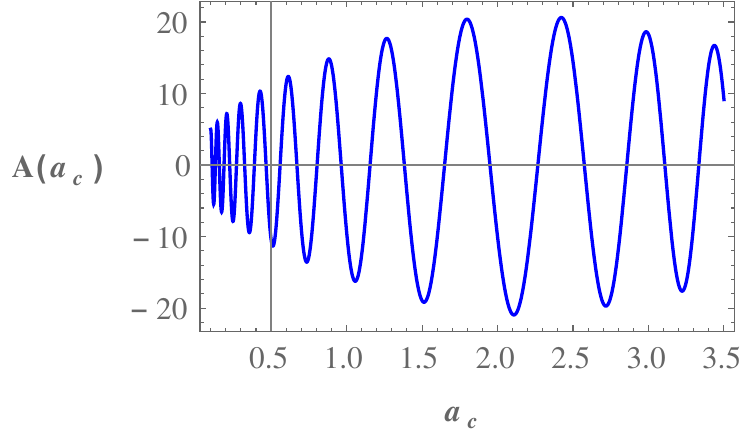}
\hspace{0.1cm}
\centering\includegraphics[width=2.5in]{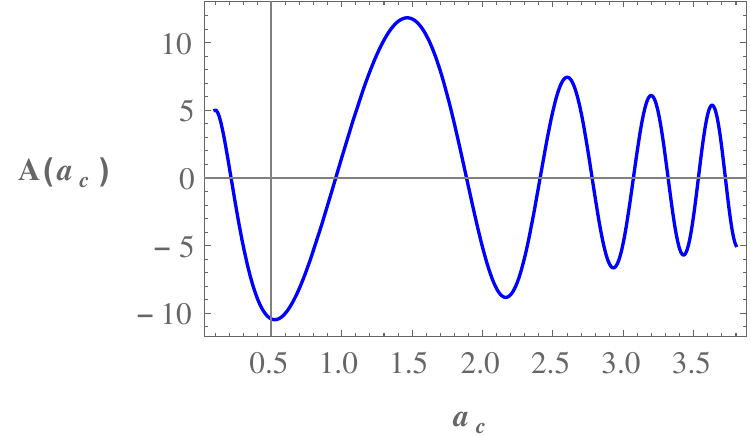}
\caption{The he behavior of the $A(a_c)$ (see Eq. \eqref{NC-WDW-3}) for the
commutative case (the left panel) and NC model (the right panel).
We have assumed $\theta=-0.5$, $V_0=0.4$, ${\cal W}=2$, $D=4$, $\Pi(0)=10$, $a(0)=18$, $A(0.1)=5=A'(0.1)$.
}
\label{Dep-IC}
\end{figure}

Our numerical approach investigations have shown that the
solutions of equation \eqref{NC-WDW-3} is very sensitive to the initial conditions and model parameters, as expected.
Meanwhile, it is worth noting that the choice of the eigenvalue $m$ is
dependent on the other CIs, thus we cannot consider the same value for
it in the commutative and NC models except in one particular case.
 More specifically, the eigenvalue $m$ must be determined from Equation \eqref{con-of-motion}
 assuming the classical ICs for ${\Pi}_{\phi}(0)$ and $a(0)$ which are employed to generate
 the solutions for equations \eqref{diff.eq1}-\eqref{diff.eq4}.
 The only scenario in which the eigenvalue $m$ is the same for both commutative and NC
 instances is when $a(0)=0$. However, ignoring this particular case, we always get
 different values for $m$ in the commutative and NC models. For example, assuming
 $\Pi_{phi}(0) =10$, $a(0)=18$, and $\theta=-0.5$, we get $m=10$ for the commutative
 case and $m=1$ for the NC case. Figure \ref{Dep-IC} depicted this case.
 It is worth noting that adopting an NC scenario with an NC parameter of
 order one is consistent with the fundamental quantum gravity scale.
In the next section, we present some comments regarding our herein NC quantum case.

\section{Summary and discussions}
\label{Concl}

In this paper, we considered the FLRW metric and an extended version of the standard SB theory as
background geometry and underlying framework in arbitrary dimensions.
By introducing a dynamical deformation in the phase space and using the
Hamiltonian formalism, the NC field equations were obtained, which reduce
to their corresponding commutative  counterparts  when the NC parameter vanishes.

Our main goal was to study the evolution of the universe
at early times in a (semi)classical and quantum regimes
based on commutative and NC models.
In this regard, we have tried to simplify our model while
maintaining its realism. For example, we have just considered the
spatially flat FLRW metric and assumed that the scalar field is dominant.
Furthermore, we hypothesized that the scalar potential has a specific form that is useful in the classical case.

Regarding the classical NC case, using the scalar potential \eqref{A-1}
(that leads to the power-law inflation in the standard models),
we have applied two methods to analyze the behavior of the physical quantities.
In the first method, we used the numerical analysis, which gives the
following results for the small values of the NC parameter.
(i) In the earlier times, the scale factor is accelerated, so after
a very short time it is decelerated. The later epoch can be assigned
to the universe dominated by radiation. This successful graceful exit
obtained by the presence of NC effects. (ii) With regard to the horizon
problem, we have considered the nominal condition, which is satisfied for our classic NC model.
These results indicate that our NC scenario can be considered as
a successful inflationary model. (iii) At late times, $\ddot{a}$
asymptotically approaches zero, which can be attributed to the
coarse-grained explanation by quantum gravity effects.
In Section \ref{SecIV}, by constructing a suitable configuration
for the dynamical system, we plotted the phase portrait of
equations \eqref{y-dyn-arb-n} and \eqref{y-dyn-2} that confirms the above consequences.
(iv) At the level of the field equations, we obtained an
evolutionary equation for the scale factor to compare it with
the Starobinsky model. It seems that the NC parameter can play
a crucial role in finding a correspondence between these models.

However regarding the second method of our herein NC classical model,
we took the ansatz \eqref{inf-1} as the solution of the field equations
that leads to an exact solution for the scalar field. Our numerical
investigations demonstrate that considering the solutions for the scale
factor and the scalar field associated to this method indicates that the
modified Klein--Gordon equation is still valid (or equivalently the
conservation of the EMT \eqref{cons-eq} is satisfied)
only when the NC parameter takes very small values.
In what follows, let us mention some interesting features
of these solutions. (i) The evolution of the scale factor
bears close resemblance with that obtained in pre-Big-Bang
scenarios \cite{GV03}: concretely, from \eqref{inf-1},
we obtain $\dot{H}(t)=-\frac{t}{\Theta}H(t)$. (ii) We
can easily show that the number of e-folds is given by
$\mathcal{N}=H_i\sqrt{2 \pi  \Theta },$ as
where we assumed the asymptotic values of the scale factors
$a_i$ and $a_f$ at $t=\pm\infty$. According to the corresponding
 field equations, in four dimensions, we have $H_i=(8\pi G \rho_{i}/3)^{1/2}$,
 where $\rho_{i}$ is the value of the energy density of the universe at early times.
For instance, assuming $\mathcal{N}\sim 60$, we obtain $\sqrt{\Theta}\sim 3.6 l_{pl}$,
where $l_{pl}$ denotes the Planck length. This results indicates that when only the
 scalar field dominates, the number of e-folds depends on the
 non-commutative parameter which is of order the Planck length.

In Section \ref{SecV}, we have investigated the commutative quantum
cosmology in arbitrary dimensions. We have obtained solutions for the
WDW equation analytically for the same potential as in the classical models.
The classical limit of the wave function was then found using the WKB approximation,
and it was demonstrated that these solutions are the same as those obtained in Section \ref{SecIII}.

At the NC quantum level, we focused our research on a specific but fascinating situation.
In contrast to the extremely complicated form of the NC-WDW equation, we
have shown that there is a constant of motion that commutes with the
Hamiltonian in the constraint space of states. The strategy
of exploiting this important conclusion helped transform the NC--WDW
equation into an ordinary differential equation that can be
solved numerically in the general case or analytically in
approximate cases.

In the general scenario, eigenvalue $m$ cannot have the same
value for the commutative and NC models. It is obtained in more
concrete terms by choosing the classical initial values for the
scale factor and momentum associated with the scalar field.
Our numerical research have revealed that for the NC scenario,
damping behavior in the wave function is always conceivable for
values of the NC parameter close to one. Because of the postulated
noncommutativity of the momenta, such intriguing quantum
characteristics of the very early universe were obtained.

Let us conclude this section by underlying a few essential points.

\begin{itemize}

\item

It is noteworthy that, in general, one might not consider the proper transformations
that provide the corresponding usual gravitational model with the canonical
kinetic term from the action \eqref{SB-action}.
Nonetheless, given the significance of the subject, let us consider a specific
transformation $d\tilde{\phi}= \sqrt{J(\phi)}d\phi$ that takes
the gravitational sector of the SB model \eqref{SB-action} to \cite{YZ21} (see also \cite{QH23})
\begin{eqnarray}
 {\cal S}^{^{(D)}}=\int d^{^{\,D}}\!x \sqrt{-g}\,
 \Big[R^{^{(D)}}-g^{\alpha\beta}\,({\nabla}_\alpha\tilde{\phi})
 ({\nabla}_\beta\tilde{\phi})-U(\tilde{\phi})\Big]
\label{canon-1},
\end{eqnarray}
where ${\cal W}\phi^n\equiv J(\phi)>0$ and $U[\tilde{\phi}(\phi)]=V(\phi)$.

Moreover, by expressing  $J(\phi)$  in terms of the potentials, equation \eqref{canon-1} transforms to
\begin{eqnarray}
 {\cal S}^{^{(D)}}=\int d^{^{\,D}}\!x \sqrt{-g}\,
 \Big[R^{^{(D)}}-\left(V_{,\phi}\frac{dU^{-1}(V(\phi))}{dV(\phi)}\right)^2g^{\alpha\beta}\,({\nabla}_\alpha\phi)
 ({\nabla}_\beta\phi)-V({\phi})\Big],
\label{canon-2}
\end{eqnarray}
where $U^{-1}$ denotes the inverse function of $U$.

We observe that \eqref{canon-1} and \eqref{canon-2} are equivalent actions
that could provide identical predictions \cite{YZ21}.
It is crucial to emphasize that for any non-canonical model with the specific coupling function
(see, for example, the SB model with $J(\phi)= {\cal W}\phi^n>0$) using
the previously mentioned transformation
to produce the canonical kinetic term may restrict one to take a particular canonical
potential. Furthermore, in actual practice, when a particular
cosmological model is studied, it is sometimes much simpler to
work with only one of the equivalent actions (i.e. either \eqref{SB-action} or \eqref{canon-1}) than the other.
This makes it easier to obtain analytical or numerical results
and to analyze the consequences.
With the other equivalent action, however, it can be more challenging or even impossible.

It is important to highlight that, the NC models we presented here, most notably the NC quantum model (see section \ref{SecVI}), we believe that, to the best of our knowledge, the work we have done is completely new and has nothing in common with earlier models, even when the transformations mentioned above are taken into consideration.
Moreover, in regard to subsection \ref{Second method}, we stress that our NC classical model, which is based on the simple deformed phase space algebra, produces the same results as the intriguing NC inflationary model found in Ref. \cite{R11}. This, in turn, indicates a correspondence between these gravitational models and can be applied in a variety of contexts for both model parameters comparison and structural phenomenological discussions. Furthermore, subsection \ref{First method} also represents interesting numerical consequences. Concretely,
it not only indicates an inflationary model that can solve the problems associated with the equivalent commutative models, but also corresponds with the Starobinsky model, see for instance \cite{V85}, and references therein.

Eventually, we believe that the models developed in this paper, which
consider an intriguing generalized noncommutativity, arbitrary
dimensions, and a gravitational scalar model with a non-canonical
kinetic term including additional parameters, have yielded new and interesting results, and have
the potential to represent correspondence with previous quantum gravity models.

\item

It is important to note that the noncommutativity we have chosen for this work
(in both the classical and quantum regimes) is not unique. For instance, in our model, in
addition to the noncommutativity between momenta, we can also propose a noncommutativity
between the scale factor and the scalar field. When this happens, the equations undoubtedly
become more complex, leading to greater degrees of freedom and ultimately more appropriate solutions.
We should mention that the amplitude $|\Psi (a,\phi)|^2$ may be replaced with another
quasiprobability distribution  \cite{BBDP08} that is generated using the
Moyal product in order to obtain probability measures in the NC models.

\item

Finally, it is worth noting that the NC cosmological models have both
advantages and shortcomings (some of which will be summarized in the
following paragraphs, respectively), compared to the corresponding
commutative models in the classical and quantum regimes.

NC models have been used to eliminate singularities such as the big bang, which were
anticipated by the corresponding commutative cosmological model.
These models can lead to an inflationary phase in the early universe
(without using ad hoc inflationary potentials, as in standard models) and
the quantum gravity effects more effectively than the commutative models, resulting in a
more complete description of the universe at small scales. They may also
produce novel observables and provide opportunities for observational verification.

The field equations associated with the NC models are
more complicated to solve than those of the corresponding
commutative models, making mathematical analysis and calculations
more difficult. Moreover, physical interpretation of the consequences
related to the NC models is challenging because they may add atypical
aspects that are difficult to understand. Furthermore, NC models have
not been experimentally tested, making it difficult to determine their
validity and compare them to the standard models.

\end{itemize}

\section*{Acknowledgments}
SMMR acknowledges the FCT grants UID-B-MAT/00212/2020 and UID-P-MAT/00212/2020
at CMA-UBI plus the COST Action CA18108 (Quantum gravity phenomenology in the multi-messenger approach).



\bibliographystyle{utphys}

\end{document}